\documentclass{jetpl}
\twocolumn

\lat

\usepackage{bm,url}
\usepackage{epsfig}
\usepackage{graphicx}

\def\be{\begin{equation}}
\def\ee{\end{equation}}
\def\ba{\begin{eqnarray}}
\def\ea{\end{eqnarray}}

\def\lsim{\raise0.3ex\hbox{$\;<$\kern-0.75em\raise-1.1ex\hbox{$\sim\;$}}}
\def\gsim{\raise0.3ex\hbox{$\;>$\kern-0.75em\raise-1.1ex\hbox{$\sim\;$}}}

\title{The GZK horizon and constraints on the cosmic ray source spectrum 
from observations in the GZK regime} 

\rtitle{GZK horizon and constraints on the cosmic ray source spectrum 
from observations in the GZK regime} 

\sodtitle{GZK horizon and constraints on the cosmic ray source spectrum 
from observations in the GZK regime}

\author{M.~Kachelrie\ss$^1$, E.~Parizot$^2$, and D.~V.~Semikoz$^{2,3,4}$} 

\address{$^1$~Institutt for fysikk, NTNU, N--7491 Trondheim, Norway}
\address{$^2$~APC, 10, rue Alice Domon et Leonie Duquet, F--75205
  Paris Cedex 13, France} 
\address{$^3$~CERN Theory Division, CH--1211 Geneva 23, Switzerland} 
\address{$^4$~INR RAS, 60th October Anniversary prospect 7a,
  117312 Moscow, Russia}

\abstract{
We discuss the GZK horizon of protons and present a method to constrain 
the injection spectrum of ultrahigh energy cosmic rays (UHECRs) from supposedly 
identified extragalactic sources. This method can be applied even when only 
one or two events per source are observed and is based on the analysis of the 
probability for a given source to populate different energy bins, depending 
on the actual CR injection spectral index. In particular, we show that for a
typical source density of $4\times 10^{-5}\,\mathrm{Mpc}^{-3}$, a data set
of 100 events above $6\times 10^{19}$~eV  allows one in 97\% of all cases
to distinguish a source spectrum $dN/dE\propto E^{-1.1}$ from one with 
$E^{-2.7}$ at 95\% confidence level. 
}


\begin{document}

\maketitle

{\em Introduction---}%
%
One of the main obstacles to fast progress in cosmic ray (CR) physics has 
been the impossibility to identify individual sources. However, there are 
two pieces of evidence indicating that we are at the dawn
of ``charged particle astronomy.'' First, anisotropies on medium scales
have been found combining all available data  of 
``old'' CR experiments~\cite{msc} as well as in the data from the Pierre
Auger Observatory (Auger)~\cite{ICRC}. Second, the Auger data hint for
a correlation of UHECRs and active galactic nuclei (AGN)~\cite{corr},
although this correlations has been contested~\cite{con}. 
Thus one may
anticipate that the influence of extragalactic magnetic fields 
is small so that UHECRs are not
significantly deflected from their initial direction. 
This should be particularly true above the  GZK cutoff~\cite{gzk}
at $\approx 5\times 10^{19}$eV, when the range of UHECRs is
significantly reduced by their interactions with photons from the 
cosmic microwave background (CMB). For instance, for
typical energy spectra and sources distributed roughly
homogeneously throughout the universe, 70\% of the protons with an
observed energy of 80~EeV come from sources
closer than 100~Mpc, even accounting for a 20\% error in the energy
determination. Over such distances, the angular spread caused by
random magnetic fields of 1~nG is typically $\lsim 3^\circ$ for such
high-energy protons. Deflections in the Galactic magnetic field are
expected to be of the same order of magnitude~\cite{GMF}.

The main reason why no sources have been identified yet would be in
this scenario that the accumulated sky exposure is not yet large enough. 
While larger exposures will inevitably increase the number of UHECRs
detected per source, it may take many years until
enough events are accumulated from even the most intense source in the
sky to allow one drawing a decent individual spectrum. The diffuse
energy spectrum of CRs below $E\lsim 4\times 10^{19}\,$eV is known
with reasonable accuracy and requires a generation spectrum
$dN/dE\propto E^{-\alpha}$ with $\alpha\approx 2.7$ for identical
sources---or an appropriate distribution of maximal energies
$E_{\max}$~\cite{Emax}---while both the source and the diffuse
spectra at higher energies are essentially unknown. It is therefore
timely, in the intermediate phase when sources may be
identified by correlation studies but typically only one or two events
per source are detected, to ask how the injection 
spectrum can be
determined best. 

While first-order
Fermi shock acceleration typically results in $\alpha$ around 
2.1~\cite{standard}, there 
exist various models that predict either much harder or softer spectra.
An example for a model with $\alpha \sim 1$ up to $10^{20}\,$eV is the
acceleration in the electric field around supermassive black
holes suggested in Ref.~\cite{Neronov:2003,Neronov:2004ga} that
explains also the observed properties of large scale jets in
AGN~\cite{Neronov:2002se}. Another possibility to obtain $\alpha\sim
1$ is to take into account a large photon background in the
acceleration region in the usual shock acceleration~\cite{Derishev:2003qi}.
On the other hand, pinch acceleration may serve as an example for
$\alpha=2.7$~\cite{pinch}. 

In this work, we present an alternative method to set constraints on the UHECR
source spectrum, suitable for the near future of proton astronomy. 
The basic idea to constrain the spectral index of individual sources
is that, even though the relative weight of different sources cannot
be known in advance (i.e. before measuring their spectra
individually), the relative weight of different energy bins
\emph{for a given source\/} is a direct consequence of the source
spectrum. Now suppose that a minimal energy $E_{\min}$ can be
identified, above which we can trust that the observed CRs
come roughly in straight lines from their source and, most
importantly, sources inside the horizon appear with a small enough
angular spread 
on the sky that they do not overlap. The energy distribution of 
CRs seen above $E_{\min}$ from a given source should then
reflect the source spectrum (modified by the usual propagation
effects), and even if one observes only one of them, its energy
contains  some information about the source
spectrum.  We show how this simple argument can be implemented
quantitatively for a given data set, taking into account UHECR energy
losses from pure proton sources with supposedly identified distances
and identical maximum energy. We use this toy model to illustrate the
basic features of the method and to explore its potential power, leaving
necessary refinements for future work. 

{\em Propagation and horizon scale of UHE protons---}%
%
\begin{figure}
\begin{center}
\epsfig{file=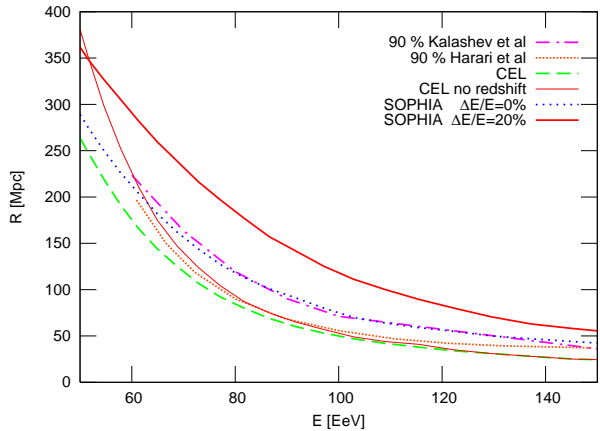,width=.33\textwidth,angle=270}
\end{center}
\caption{\label{fig:0}
Fig.~1: Distance $R$ in Mpc from which 90\%  of UHECRs arrive with energy $>E$ 
as function of the threshold energy $E$ for $E_{\max}=10^{21}\,$eV and 
$\alpha=2.7$. The thin solid red
line uses CEL in a static Universe as \cite{Harari:2006uy}, the green line 
uses CEL in an expanding
Universe. The blue line labeled ``SOPHIA'' has to be compared 
to~\cite{Kalashev:2007ph}. The red
line takes into account additionally an experimental energy resolution 
$\Delta E/E =20$ \%.}
\end{figure}
\begin{figure}
\begin{center}
\epsfig{file=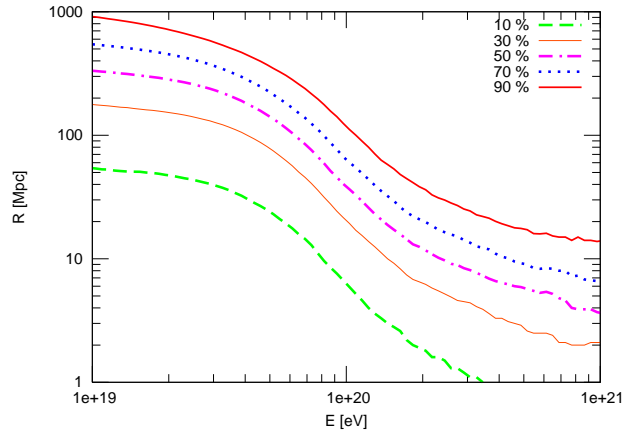,width=.33\textwidth,angle=270}
\end{center}
\caption{\label{fig:2}
Fig.~2: The distance $R$ in Mpc for which a certain fraction $f$ of UHECRs
arrives with energy $>E$ as function of the energy threshold $E$ for
$\gamma=2.7$. From top to bottom,   $f=90\%$ as red line,
$f=70\%$ as pink,  $f=50\%$ as magenta, $f=30\%$ as blue and
$f=10\%$ as green line.
}
\end{figure}
In Fig.~\ref{fig:0}, we show the ``90\% horizon'' -- i.e. the
distance $R_{90}$ from which 90\% of the UHECRs observed above a given
energy, $E$, originate---as function of energy. 
We assume a uniform source distribution with a density $n_s=4 \times
10^{-5}/{\rm Mpc}^3$ (cf.\ e.g.\ Refs.~\cite{ns,agn})
and a power-law source spectrum $dN/dE\propto E^{-\alpha}$ with
$\alpha=2.7$ up to the maximal energy $E_{\max}=10^{21}\,$eV. 
We used for the calculation of photo-pion production the program 
SOPHIA~\cite{sophia}, either taking into account the
stochasticity of the corresponding energy losses (dotted, blue line) or applying
the continuous energy loss (CEL) approximation to its results (dashed,
green line). The $e^+e^-$ pair production losses were taken from 
Ref.~\cite{ee}.

The $f=90\%$ horizon computed within the CEL approximation underestimates
considerably the full Monte Carlo result. The difference increases for
a larger ``horizon fraction'', $f\to 1$, and as function of energy
for $E\to E_{\max}$.  There are two 
reasons for the latter discrepancy. First, the energy transfer
per interaction, $y$, increases with energy and violates more and more
strongly the formal requirement $y\ll 1$ needed for the applicability
of the CEL approximation. Second, the flux taking into account the
stochastic nature of the energy losses in pion production remains
finite for $E\to E_{\max}$, while in the CEL approximation no
particles with $E=E_{\max}$ can reach the observer from a source at a
finite distance~\cite{CEL}. 

In a realistic experiment, the primary energy can
only be reconstructed with a finite precision. Assuming a Gaussian (in
$\log E$) experimental uncertainty of $\Delta E/E = 20$\%, we computed
the 90\% horizon as a function of the \emph{measured\/} CR energy, for
the same conditions as above. The two resulting curves are also shown
in Fig.~\ref{fig:0}. Since the CR spectrum is falling steeply, the
misinterpretation of lower energy events as high energy ones has a
larger impact than the reverse, which in turn leads to an increase of
the estimated horizon scale.  At low energies, say $\lsim 5\times
10^{19}\,$eV, the observed spectrum approximates well to a power-law
and the energy resolution only affects the absolute flux, not the
relative fluxes relevant for $R_{90}(E)$.

The horizon scale for UHE protons and nuclei was recently discussed
also in Refs.~\cite{Harari:2006uy,Kalashev:2007ph}. In 
Fig.~\ref{fig:0} we compare our calculations to those of
Refs.~\cite{Harari:2006uy} and \cite{Kalashev:2007ph} for proton
primaries. In Ref.~\cite{Harari:2006uy}, Harari {\it et al.\/} 
presented results (shown as orange line) using the CEL
approximation and assuming a static Universe, our result for the same
assumptions is shown with a thin solid red line. 
Both calculations agree well at moderate energies $E=80-100$ EeV,
while there is some disagreement both at high and low energies. 
However, the differences at low energies between the two calculations
are  much smaller  than the differences between those calculations and
the more correct CEL calculation in the $\Lambda$CDM model for the expanding 
Universe, presented with a green line.

All results using the CEL approximation differ in shape as a function of energy 
from the calculationss using SOPHIA for pion production 
either directly (blue line), or using the SOPHIA results in a kinetic equation
approach as in Ref.~\cite{Kalashev:2007ph} (magenta line). The
agreement between the latter two results is almost perfect at all energies.

As an illustration, we show in Fig.~\ref{fig:2} the horizon distance
corresponding to different CR fractions. Specifically, we plot the
distance $R_{f}$ below which a given fraction $f$ of the UHECRs reach
the Earth with an energy larger than $E$, as a function of that energy, for
$f = 10$\%, 30\%, 50\%, 70\% and 90\% (using always SOPHIA and $\Delta
E/E=20$\%).

{\em Estimation of the spectral index---}%
%
Since the angular resolution of cosmic ray experiments is poor by
astronomical standards, the identification of individual sources
requires a relatively large angular distance between them. This can
only hold for sufficiently high energies such that the horizon scale
is small, say of the order of 100~Mpc, leaving a limited number of
sources over the sky. Defining as 
horizon, within which 90\% of all CRs observed above a given energy
were emitted, we find from Fig.~\ref{fig:2} that a horizon of 100~Mpc
corresponds to a threshold energy of $E = 1\times 10^{20}\,$eV. At
present, the importance of deflections in extragalactic magnetic
fields above this energy is unclear. As soon as sources are detected,
one will be able to set an upper limit and to a certain extent
reconstruct the extragalactic magnetic field. Here, we limit ourselves
to the optimistic scenario where deflections in extragalactic magnetic
fields are not much larger than the combined effects of the Galactic
magnetic field and the experimental angular resolution. 

%
At present, the picture of uniformly distributed, extragalactic UHECR
sources having all the same luminosity and the same injection spectrum
is able to describe well the observed energy spectrum in a broad
energy range from a few$\times 10^{17}$~eV or a few$\times 10^{18}$~eV 
up to the GZK cutoff, depending on the assumed source
composition~\cite{dip,nuc}. 

We first produce a Monte Carlo (MC) sample by generating sources with
constant comoving density $n_s=4 \times 10^{-5} {\rm Mpc}^{-3}$ up to a 
maximal redshift of $z=0.1$. Then we choose a source $i$ 
according to the declination dependent exposure of Auger, 
with an additional weight chosen according to
the source distance. Finally, we generate a CR with an initial
energy drawn randomly according to the assumed injection spectrum,
$dN/dE\propto E^{-\alpha_0}$, and  propagate it until it either
reaches the Earth distance or loses energy down to below $E_{\min}$.
In the former
case, we then apply an energy-dependent angular deflection to mimic
the effect of the Galactic magnetic field, with a shift perpendicular
to the Galactic plane equal to $\delta b  = 2^\circ (E/10^{20} {\rm
  eV})^{-1}$, where this magnitude is motivated by the results of
Ref.~\cite{GMF}. The chosen magnetic field likely overestimates 
deflections far away from the galactic plane in most of models. However, we 
consider this choice as a conservative upper limit.
Finally, we deflect the CR direction to account for a
finite experimental angular resolution, taking the Auger surface
detector as a reference~\cite{resolution}, with a spherical Gaussian
density $\propto \exp(-\ell^2/(2\sigma_l^2))\sin(\ell)d\ell$, where
$\sigma_\ell = 0.85^\circ$ and $\ell$ is the angular distance. 

After having generated $N$ cosmic rays, we perform a correlation
analysis between the CRs and the sources. First, we identify as ``the
source'' of a given CR the source with the smallest angular distance
$\ell$ to the observed CR arrival direction and maximal distance
$R=100\,$Mpc. Inside this region, there are around $\sim 160$ sources
for chosen density $n_s=4 \times 10^{-5} {\rm Mpc}^{-3}$. Such a small
number makes the probability negligible that sources overlap, if they
are uniformly distributed. This probability increases, if sources
follow---as expected---the large-scale structure of matter and may
constitute a real limitation to resolve single sources in cluster cores.

 Additionally, we require that the angular distance
$\ell$ be smaller than a prescribed value, $\ell_{\max}$. Next, having
pre-defined an energy $E_{2}$ that divides the whole energy range into
two large bins, we count for each source $i$ the numbers $N_{i,1}$ and
$N_{i,2}$ of high energy ($E \ge E_2$) and low energy events
($E_{\min} \le E < E_2$), respectively. Given the corresponding
fractions $f_1(\alpha)$ and $f_2(\alpha)=1-f_1(\alpha)$ of
$N_i=N_{i,1}+N_{i,2}$ events expected from a source at the identified
distance for an arbitrary value of the spectral index $\alpha$, we
calculate with a binomial distribution the probability, 
\be
 p_i(N_{i,1},N_{i,2}|\alpha) =
 \frac{(N_{i,1}+N_{i,2})!}{N_{i,1}!N_{i,2}!} f_1^{N_{i,1}}(\alpha)
 f_2^{N_{i,2}}(\alpha), 
\ee
that the observed numbers $N_{i,j}$ are consistent with the value
$\alpha_0$ used in the MC. 
Considered as a function of $\alpha$, this probability distribution
has the true value $\alpha_0$ as its expectation value, if our
procedure is unbiased, and measures
how strongly the data disfavor a differently assumed value
$\alpha\neq\alpha_0$.  
 
Since the different sources emit CRs independently from one another,
we can simply multiply the single source probabilities
$p_i(N_{i,1},N_{i,2}|\alpha)$ to obtain the global probability of a
given data set with $N_{\mathrm{s}}$ identified sources: 
\be
p(\{N_{i,1}\}_{i=1,N_{\mathrm{s}}}|\alpha) =
\prod_{i=1}^{N_{\mathrm{s}}} p_i(f_{i,1},f_{i,2}|\alpha). 
\label{prob_MC}
\ee

The basic outcome of a sample of MC simulations for fixed
parameters $\theta=\alpha_0\ldots$ is thus a binned distribution,
$f(p|\theta)$, giving the fraction $f$ of MCs producing the value $p$.  
With how much confidence can we distinguish these distributions for
two different $\theta_1$ and $\theta_2$? Clearly, the smaller the
overlap of the two distributions, the easier the two parameter sets
$\theta_{i}$ can be distinguished.

%
\begin{figure}[h]
\begin{center}
\epsfig{file=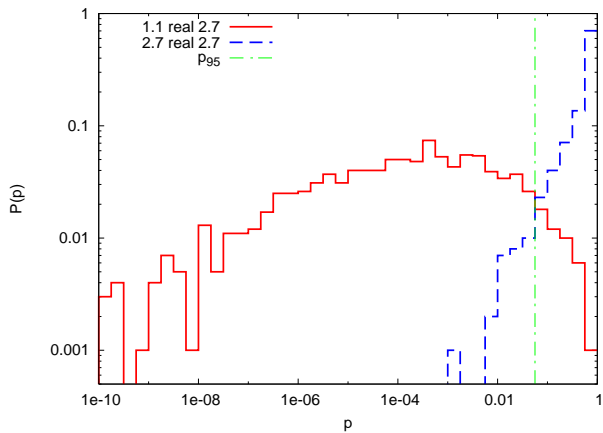,width=.33\textwidth,angle=270}\\
\end{center}
\caption{\label{fig:3} 
Fig.~3: Distribution of probability of reconstructed power 
law spectrum if real power law spectrum is 
$\alpha=2.7$, angle $\ell_{\max}=4^\circ$. In all cases $E_{\min} = 60$ EeV. 
The red line
is for $\alpha=1.1$ and the blue line for $\alpha=2.7$.}
\end{figure}
We study now the possibility to distinguish different values of the 
injection spectrum of CRs in more detail. As simplifying assumption we assume
that the injection spectrum of all sources is the same, i.e.\ in
particular that the maximal energy of all sources is identical.
This assumption allows us to study the spectra only above
$\sim 4\times 10^{19}\,$eV, because at lower energies a spectral
index  $\alpha<2.6$ requires either additional Galactic sources or 
a non-uniform source distribution. In the latter case, either the
source density or the luminosity of single sources should increase as
function of redshift, $n(z)=n_0(1+z)^m$ and  $L(z)=L_0(1+z)^m$ 
respectively,  or  the maximal energy of sources is
distributed as $dn/dE_{\max}\propto E_{\max}^{3.6-\alpha}$~\cite{Emax}.  
Moreover, we consider only two extreme cases, namely a power-law with
$\alpha_0=1.1$ and $\alpha_0=2.7$. 

In Fig.~\ref{fig:3} we compare the distributions of probabilities obtained
from Eq.~(\ref{prob_MC}) choosing as true value $\alpha_0=2.7$,
as source density as always $n_s=4\times 10^{-5}/{\rm Mpc}^3$, 
as number of CRs $N=100$, and $\ell_{\max}=4^\circ$.
The red solid line is the distribution of probabilities obtained
assuming $\alpha=1.1$, while the blue dashed line corresponds
$\alpha=2.7$. The two curves have only a small overlap, since the
probabilities using the correct $\alpha$ are rather narrowly
concentrated around $p=1$, while the probability distribution using
the wrong $\alpha$ extends from extremely low values up to one.
Thus an experimental differentiation between different injection
spectra seems possible, even if only one or, in few cases, two
events per source are detected, as it is the case for the chosen
parameters in Fig.~\ref{fig:3}. This constitutes the main result of
our work. 

We quantify the chances to distinguish two different spectral indices
in the following way: We calculate the area $A$ corresponding to the
desired confidence level (C.L.), $A$, starting from 1 to the left using
the best-fit distribution (e.g.\ the blue line in Fig.~\ref{fig:3}) 
and obtain thereby as its lower boundary $p_A$. Thus only in $1-p_A$ cases
we will obtain by chance a lower probability using the correct test
hypothesis. Next we count how large is the area $B$ of the wrong test
hypothesis on the left of $p_A$. As final answer we obtain that in the
fraction $B$ of all cases we can distinguish between the two hypotheses
with C.L. $A$.  

Let us illustrate this procedure for the case considered above, 
choosing as confidence level $A=95\%$. The green dashed-dotted vertical line in
Fig.~\ref{fig:3} enclosing 95\% of the area of the true (blue)
distribution determines $p_{95}=0.056$.  The area of the red curve on
the left of $p_{95}=0.056$ is $B=0.971$. Hence one can exclude in
$B=97.1\%$ of cases with at least 95\% C.L. the exponent $\alpha=1.1$ for 
the spectrum, if the true exponent is $\alpha_0=2.7$.

In addition to the rather extreme cases of the spectral indices above, 
we investigated the ability of the method to distinguish between any of 
them and an intermediate value of $\alpha_{0} = 2.0$, often considered in 
the context of astrophysical particle acceleration. As an illustration, 
we found that with a data set of 100 cosmic rays above $6\times 10^{19}$~eV, 
it is possible in 50\% of the cases to discriminate $\alpha_{0} = 2.0$ 
from a value of either 1.1 or 2.7 with a C.L. of 95\%. Likewise, for a 
data set of 200 cosmic rays above $4\times 10^{19}$~eV (i.e. for essentially 
the same exposure of the sky, but with a lower energy threshold), an 
injection spectral index of 1.1 can be discriminated against $\alpha_{0}=2.0$ 
with a C.L. of 95\% in 90\% of the cases, while an injection spectral index 
of 2.7 can be discriminated against $\alpha_{0} = 2.0$ with a C.L. of 95\% 
in 70\% of the cases.

{\em Summary---}%
%
We have proposed a method to estimate the generation spectrum of
individual extragalactic CR sources that is well-suited for the time 
when only one or
two events per source are detected. An important ingredient of this
method is the relative fraction of events contained in a prescribed 
energy interval. Therefore we have recalculated the horizon scale of
ultra-high energy protons, taking into account a reasonable energy
resolution, similar to that of Auger.  

We have demonstrated for a toy-model the potential of this method,
finding that around 100 events above $6\times 10^{19}\,$eV are
required to distinguish with 97\% probability at least at the 95\%
C.L. the two extreme cases $\alpha=1.1$ and 2.7. A differentiation
between  $\alpha$'s that are more similar will be clearly more
challenging. An injection spectral index of 2.0 can still be distinguished 
from the two above values with a 95\% C.L. in the majority of cases 
(with the same statistics).

Several of the issues
we have neglected, like the effect of a possible $E_{\max}$
distribution, should be included in a more complete study as soon as
experimental data will be available.  A proper estimation of 
$\alpha$ also requires to quantify the bias introduced e.g.\ by
misidentified events. In general, it proves more efficient to remove from 
the data set the doubtful events (e.g. in regions where a given catalogue 
used to identify sources is known to be incomplete, or when several sources 
at different distances are identified over a small region of the sky, with 
possible overlap due to magnetic deflection or poor angular resolution), 
and apply the method with a correspondingly smaller statistics. Sources 
physically clustered in the universe are not a problem here, since they 
are located essentially at the same distance from the Earth and thus 
suffer from the same attenuation during propagation.

\section*{References}


\begin{thebibliography}{99}

\bibitem{msc}
M.~Kachelrie{\ss} and D.~V.~Semikoz,
Astropart.\ Phys.\  {\bf 26}, 10 (2006)
[astro-ph/0512498].

\bibitem{ICRC}
S.~Mollerach {\it et al.}, 
to appear in
{\em Proc. 
Mexico, 2007\/},
arXiv:0706.1749 [astro-ph].

\bibitem{corr}
J.~Abraham {\it et al.}  
Astropart.\ Phys.\  {\bf 29}, 188 (2008)
[astro-ph/0712.2843].

\bibitem{con}
R.~U.~Abbasi {\it et al.} [HiRes Collaboration],,
arXiv:0804.0382 [astro-ph];
D.~S.~Gorbunov {\it et al.}, 
  arXiv:0804.1088 [astro-ph];
I.~V.~Moskalenko {\it et al.}, 
  arXiv:0805.1260 [astro-ph].

\bibitem{gzk}
K.~Greisen,
Phys.\ Rev.\ Lett.\  {\bf 16}, 748  (1966);
G.~T.~Zatsepin and V.~A.~Kuzmin,
JETP Lett.\  {\bf 4}, 78 (1966)
[Pisma Zh.\ Eksp.\ Teor.\ Fiz.\  {\bf 4}, 114 (1966)].

\bibitem{GMF}
M.~Kachelrie\ss, P.~D.~Serpico and M.~Teshima,
Astropart.\ Phys.\  {\bf 26}, 378 (2006)
[astro-ph/0510444].

\bibitem{Emax}
M.~Kachelrie{\ss} and D.~V.~Semikoz,
Phys.\ Lett.\  B {\bf 634}, 143 (2006)
[astro-ph/0510188].


\bibitem{standard}
V.~S.~Berezinskii {\it et al},
Astrophysics of cosmic rays,
Amsterdam: North-Holland 1990.
T.~Gaisser, 
Cosmic Rays and Particle Physics,
Cambridge University Press 1991.
R.~J.~Protheroe and R.~W.~Clay,
Publ.\ Astron.\ Soc.\ of Australia  {\bf 21}, 1 (2004)
[astro-ph/0311466].

\bibitem{Neronov:2003}
 A.~Neronov and D.~Semikoz, 
New Astronomy Reviews, {\bf 47}, 693 (2003), 
A.~Neronov, D.~Semikoz and I.~Tkachev, 
arXiv:0712.1737 [astro-ph]

\bibitem{Neronov:2004ga}
A.~Neronov, P.~Tinyakov and I.~Tkachev,
J.\ Exp.\ Theor.\ Phys.\  {\bf 100}, 656 (2005)
[Zh.\ Eksp.\ Teor.\ Fiz.\  {\bf 100}, 744 (2005)]
[astro-ph/0402132].

\bibitem{Neronov:2002se}
A.~Neronov, D.~Semikoz, F.~Aharonian and O.~Kalashev,
Phys.\ Rev.\ Lett.\  {\bf 89}, 051101 (2002)
[astro-ph/0201410].

\bibitem{Derishev:2003qi}
E.~V.~Derishev {\it et al.}, 
Phys.\ Rev.\  D {\bf 68}, 043003 (2003)
[astro-ph/0301263].

\bibitem{pinch}
V.~V.~Vlasov, S.~K.~Zhdanov and B.~A.Trubnikov,
Fiz.\ Plasmy {\bf 16}, 1457 (1990).

\bibitem{ns} 
P.~Blasi and D.~de Marco,
Astropart.\ Phys.\  {\bf 20}, 559 (2004)
[astro-ph/0307067];
M.~Kachelrie{\ss} and D.~Semikoz,
Astropart.\ Phys.\  {\bf 23}, 486 (2005)
[astro-ph/0405258].

\bibitem{agn}
A.~Cuoco {\it et al.},
  Astrophys.\ J.\  {\bf 676}, 807 (2008)
  [0709.2712 [astro-ph]].


\bibitem{sophia}
A.~M\"ucke {\it et al.}, 
Comput.\ Phys.\ Commun.\  {\bf 124}, 290 (2000)
[astro-ph/9903478].

\bibitem{ee}
V.~Berezinsky, A.~Z.~Gazizov and S.~I.~Grigorieva,
Phys.\ Rev.\  D {\bf 74}, 043005 (2006)
[hep-ph/0204357].

\bibitem{CEL}
See also Ref.~\cite{ee} and 
V.~Berezinsky, A.~Gazizov and M.~Kachelrie\ss,
Phys.\ Rev.\ Lett.\  {\bf 97}, 231101 (2006)
[astro-ph/0612247].

\bibitem{Harari:2006uy}
D.~Harari, S.~Mollerach and E.~Roulet,
JCAP {\bf 0611}, 012 (2006)
[astro-ph/0609294].


\bibitem{Kalashev:2007ph}
O.~E.~Kalashev {\it et al.}, 
arXiv:0710.1382 [astro-ph].

\bibitem{dip}
V.~Berezinsky, A.~Z.~Gazizov and S.~I.~Grigorieva,
astro-ph/0210095;
Phys.\ Lett.\ B {\bf 612} (2005) 147
[astro-ph/0502550].

\bibitem{nuc}
D.~Allard, E.~Parizot and A.~V.~Olinto,
Astropart.\ Phys.\  {\bf 27}, 61 (2007)
[astro-ph/0512345].

\bibitem{resolution}
M. Ave {\it et al.}, 
in {\em Proc. ``30th International Cosmic Ray Conference'', M\'erida, Mexico, 2007\/},
\#0297.


\end{thebibliography}
\end{document}